\documentclass[amsmath, amsfonts, superscriptaddress, twocolumn,prb]{revtex4}
\usepackage{graphicx}
\usepackage{epsfig}
\usepackage{bm}
\usepackage{dcolumn}
\usepackage{amsmath}
\usepackage{amssymb}
\usepackage{color}
\usepackage{stackrel}
\usepackage{accents}
\usepackage{latexsym}
\usepackage{verbatim}
\usepackage{hyperref}
\usepackage{bbm}

\newcommand{\beg}{\begin{equation}}
\newcommand{\en}{\end{equation}}

\newcommand{\bq}{\mathbf q}
\newcommand{\bk}{\mathbf k}
\newcommand{\br}{\mathbf r}

\newcommand{\bR}{\mathbf R}

\newcommand{\veps}{\varepsilon}
\newcommand{\eps}{\epsilon}

\newcommand{\up}{\uparrow}
\newcommand{\dn}{\downarrow}
\newcommand{\dg}{^\dagger}

\begin{document}

\title{Superconductivity in Ce-based cage compounds}

\author{Suman Raj Panday}
\author{Maxim Dzero}
\affiliation{Department of Physics, Kent State University, Kent, Ohio 44242, USA}

\begin{abstract}
Cerium-based ternary compounds CeNi$_2$Cd$_{20}$ and CePd$_2$Cd$_{20}$ do not exhibit  long-range order down to millikelvin temperature range. Given the large separation between Ce ions which significantly reduces the super-exchange interactions and vanishingly small RKKY interaction, here we show that nodal superconductivity mediated by the valence fluctuations must be a  ground state in these materials. We propose that the critical temperature for the superconducting transition can be significantly increased by applying hydrostatic pressure.  
We employ an extended periodic Anderson lattice model which includes the long-range Coulomb interactions between the itinerant electrons as well as the local Coulomb interaction between the predominantly localized and itinerant electrons to compute a critical temperature of the superconducting transition. Using the slave-boson approach we show that  fluctuations mediated by the repulsive electron-electron interactions lead to the emergence of $d$-wave superconductivity. 
\end{abstract}

%\pacs{71.27.+a, 75.20.Hr, 74.50.+r}

\maketitle

\section{Introduction}
Ternary compounds CeNi$_2$Cd$_{20}$ and CePd$_2$Cd$_{20}$ are members of a family of compounds with chemical formula RT$_2$X$_{20}$ (R$=$ rare-earth element, T$=$transition-metal element and X=Al,Zn,Cd) and cubic lattice structure.\cite{Burnett2014,Yazici2015,Onimaru2010,Niemann1995,Kangas2012,Isikawa2013,Swatek2013} It was reported 
recently that no long-range order has been observed in CeNi$_2$Cd$_{20}$ and CePd$_2$Cd$_{20}$ down to temperatures in the millikelvin range even though the well formed cerium magnetic moments were observed in magnetization measurements. \cite{White2015} This surprising experimental fact may be understood by taking into account that \emph{(i)} there is a fairly large separation between the neighboring cerium ions $\sim 6.8\AA$, so that super-exchange interactions \cite{AndersonSuper} are vanishingly small and \emph{(ii)} the RKKY \cite{Ruderman1954,Kasuya1956,Yosida1957} interaction between the cerium local moments is essentially zero due to the symmetry of the lowest lying $f$-orbital multiplet.\cite{Dzero2022} Furthermore, transport measurements indicate the weak hybridization between the conduction and predominantly localized $f$-electrons. 

Vanishingly small RKKY interaction in CeNi$_2$Cd$_{20}$ and CePd$_2$Cd$_{20}$ makes these compounds analogous to CeAl$_2$. The latter, however, develops long-range antiferromagnetic order driven by the super-exchange interactions. \cite{MapleCeAl2} Therefore, in the absence of interactions which would promote magnetic long-range order,  these materials should be expected to develop superconductivity upon further cooling. Superconductivity may be driven either by electron-phonon interactions \cite{LPG2005} or by purely electron-electron interactions. \cite{ChubukovMaiti} 
It is indeed very well known by now that valence fluctuations originating from the hybridization between the conduction and $f$-electrons may lead to a superconducting instability. \cite{Lavagna1987,Read1988,Miyake2000} 

As it follows from the results of the transport and thermodynamic measurements,  hybridization between the itinerant and $f$-electrons remains weak down to low temperatures as manifested by the absence of the coherence peak in resistivity as well as low values of the Sommerfeld coefficient $\gamma$.\cite{Coleman2007} On the other hand, it is known that applying pressure will promote the valence fluctuations and, as a result, will lead to an increase in the hybridization amplitude in the $f^0\leftrightarrow f^1$ channel. As a consequence, average occupation on the $f$-site may become lower than one and electron-electron correlations may produce superconducting instability in the $d$-wave channel. 

In this paper we propose that upon further cooling CeNi$_2$Cd$_{20}$ and CePd$_2$Cd$_{20}$ will develop superconducting instability with the $d$-wave of the order parameter. In the limit of the weak coupling, the energy scale which determines the critical temperature is given by the Kondo coherence temperature, $T_{\textrm{coh}}$. One of the consequences is that applying external pressure, which promotes the fluctuations between the cerium $f^0$ and $f^1$ valence configurations, will boost $T_{\textrm{coh}}$ and superconducting transition temperature will also increase.
In this regard, these compounds may be similar to another Ce-based ternary compound CeCu$_2$Si$_2$ where similar mechanism for superconductivity was proposed awhile ago.\cite{Miyake2000} Although conceptually similar to the previous works \cite{Lavagna1987,Read1988,Miyake2000}, our work is different from the previous ones in two aspects: \emph{(i)} we use the large-$N$ approach based on the generators of the SP($2N$) group, which preserves the time-reversal symmetry and allows us to consider spin-triplet superconducting instability and \emph{(ii)} we take into account long-range Coulomb interactions between the conduction electrons. In this context, we are interested to check how the fluctuations associated with the plasmon field would affect the Cooper pairing in the nodal superconductor with repulsive interactions. \cite{Joerg1} 

In what follows, we will analyze the superconducting instability induced by the fluctuations of the bosonic fields associated with the long-rangle electron-electron correlations. Our results show that the presence of the fluctuations associated with the plasmon field leads to a significant (factor of $\sim 2$) suppression of the critical temperature, when the local $f-c$ Coulomb interactions are relatively weak. 
We find that the maximum of the critical temperature is attained in a mixed-valent regime when the average occupation number for the $f$-electrons $\sim0.8$. In this regard, these systems may provide a clearest example of superconductivity induced by strong electron-electron correlations without requiring system's proximity to a quantum critical phase transition. Unless pointed out otherwise, throughout this paper we will adopt the energy units $e=\hbar=c=k_B=1$.

\section{Model and Basic equations}
We consider a system of itinerant ($c$) and flat-band ($f$) electrons described by the following Hamiltonian:
\beg\label{Eq1}
{\cal H}=H_c+H_f+H_V+H_{fc}+{\cal H}_{C}.
\en
Here the first two terms on the right hand side are
\beg\label{HcHf}
\begin{split}
H_c&=\frac{1}{\sqrt{{\cal N}_L}}\sum\limits_{\bk\sigma}\eps_\bk c_{\bk\sigma}\dg c_{\bk\sigma}, \\
H_f&=\veps_{f0}\sum\limits_{j\sigma}f_{j\sigma}\dg f_{j\sigma}+U_{\textrm{ff}}\sum\limits_jn_{j\up}^f n_{j\dn}^f,
\end{split}
\en 
where the summation is performed over the $f$-sites, ${\cal N}_L$ is a number of lattice sites, $\eps_\bk$ is the single-particle dispersion for the itinerant electrons (to be specified below), $\veps_{f0}$ is the single particle energy for the $f$-electrons and $n_{j\sigma}^f=f_{j\sigma}\dg f_{j\sigma}$. The third term (\ref{Eq1}) accounts for the hybridization between the itinerant and $f$-electrons:
\beg\label{Hybr}
H_V=\frac{1}{\sqrt{{\cal N}_L}}\sum\limits_{\bk\sigma}\left(V_{\bk}f_{\bk\sigma}\dg c_{\bk\sigma}+V_{\bk}^*c_{\bk\sigma}\dg f_{\bk\sigma}\right),
\en
where $V_{ij}=(1/\sqrt{\cal N}_L)\sum_\bk V_\bk e^{i\bk(\bR_i-\bR_j)}$ is the hybridization amplitude. Although $V_{\bk}$ is anisotropic due to the difference in symmetry of the conduction and $f$-orbitals,\cite{Dzero2016} in what follows without loss of generality we ignore its momentum dependence $V_\bk\to V$. Due to the fact that number of the conduction and $f$-electrons are not separately conserved,  both $\veps_\bk$ and $\veps_{f0}$ are taken relative to the chemical potential $\mu$, which will be computed self-consistently. 

Lastly, the remaining last two terms in (\ref{Eq1}) describe the Coulomb interactions between the electrons:
\beg\label{HCoulomb}
\begin{split}
H_{fc}&=U_{\textrm{fc}}\sum\limits_{j\sigma\sigma'}\int d\br {\psi}_\sigma\dg(\br)\psi_\sigma(\br) n_{j\sigma'}^f \delta(\br-\bR_j),\\
{\cal H}_{C}&=\frac{1}{2}\int d\br\int d\br'{\rho}_{\textrm{c}}(\br)U(\br-\br'){\rho}_{\textrm{c}}(\br').
\end{split}
\en
Here ${\rho}_{\textrm{c}}(\br)=\sum_\sigma{\psi}_\sigma\dg(\br)\psi_\sigma(\br)$ is the density operator, $U(\br)=e^2/|\br|$ is the bare Coulomb potential and
\beg\label{Fourier}
\psi_\sigma(\br)=\frac{1}{\sqrt{{\cal N}_L}}\sum\limits_{\bq}c_{\bq\sigma}e^{i\bq\br}.
\en
 
To generate the large-$N$ expansion, we first extend the number of spin and orbital degrees of freedom for both conduction and $f$-electrons from $2$ to $N$ using the generators of the SP($N$) ($N=2k$, $k=1,2,...$) subgroup of SU($N$) to preserve the invariance with respect to the time-reversal symmetry.\cite{Flint2007}  Since the interaction between the localized $f$-electrons is assumed to be the largest energy scale of the problem, we are going to adopt the limit
\beg\label{Uff8}
U_{\textrm{ff}}=\infty,
\en
which means that we are projecting out the doubly occupied states and therefore we need to introduce slave-boson projection operators according to $f_{j\alpha}\to f_{j\alpha}b_j\dg$, $f_{j\alpha}\dg\to f_{j\alpha}\dg b_j$ ($\alpha=\pm1, ..., \pm k$) along with the constraint condition 
\beg\label{Constraint}
Q_j=\sum\limits_{\alpha}f_{j\alpha}\dg f_{j\alpha}+b_j\dg b_j=1.
\en

We can now follow the avenue of Ref. [\onlinecite{Miyake2000}].
The large-$N$ expansion is generated by rescaling $Q_j\to qN$, $b_j\to b_j\sqrt{N}$, $V\to V/\sqrt{N}$, $U_{\textrm{fc}}\to U_{\textrm{fc}}/\sqrt{N}$
and $U(\br)\to U(\br)/N$.  First, we use the path integral approach within the Matsubara formalism, so that the partition function is given by
\beg\label{PartZ}
{\cal Z}=\int{\cal D}[c c\dg f f\dg \rho \lambda]e^{-S},
\en
where $\rho$ is a real bosonic field which appears as a result of the gauge transformation $b_j(\tau)=\rho_j(\tau)e^{i\theta_j(\tau)}$, $f_{j\alpha}\to f_{j\alpha}e^{i\theta_j(\tau)}$ and $\lambda_j(\tau)\to \lambda_j(\tau)+\theta_j(\tau)$ is the slave field which is used to enforce the constraint (\ref{Constraint}). 

We employ the Hubbard-Stratonovich transformation 
\beg\label{UsefulExpressions}
\begin{split}
&\int {\cal D}\left[\sigma,\overline{\sigma}\right]\exp\left(-\frac{1}{J}\int\limits_0^\beta d\tau\left|\sigma(\tau)+J\sum\limits_{\bk \alpha} f_{\bk \alpha}\dg c_{\bk \alpha}\right|^2\right)\\&=\textrm{const.}
\end{split}
\en
to decouple the interaction terms (\ref{HCoulomb}) in the action (\ref{PartZ}) by using the bosonic fields $\varphi_{\textrm{c}}(\br,\tau)$, $\varphi_{\textrm{f}}(\br,\tau)$ and $\phi(\br,\tau)$. After this step, one can formally integrate out the fermionic fields which yields the purely bosonic action 
\beg\label{BosonicAction}
\begin{split}
S&=-N\textrm{Tr}\log\hat{G}+{N}\sum\limits_{kk'}\rho(-k)(i\lambda(k-k'))\rho(k')\\&-\frac{NU_{\textrm{fc}}}{4}\sum\limits_k\varphi_{\textrm{f}}(k)\varphi_{\textrm{c}}(-k)+\frac{N}{2}\sum\limits_k\frac{\phi(-k)\phi(k)}{U(k)}\\&-qN\sqrt{{\cal N}_L}\int\limits_0^\beta d\tau \left(i\lambda(\bk=0,\tau)\right).
\end{split}
\en
Here $q=1/N$, $U(k)=4\pi e^2/\bk^2$ and $\sum\limits_k\{...\}=T\sum\limits_{i\nu_l}(1/\sqrt{{\cal N}_L})\sum\limits_\bk\{...\}$. The first term in (\ref{BosonicAction}) is  a matrix representing a single-particle fermionic propagator:
\begin{widetext}
\beg\label{PropG}
\begin{split}
\hat{G}(k,k')=\left[
\begin{matrix}
(-i\omega_n+\eps_\bk)\delta_{\bk\bk'}+\frac{U_{\textrm{fc}}}{2}\varphi_{\textrm{f}}(k-k') + i\phi(k-k') & V\rho(k-k') \\ V\rho(k'-k) & 
(-i\omega_n+\veps_{f0})\delta_{\bk\bk'}+\frac{U_{\textrm{fc}}}{2}\varphi_{\textrm{c}}(k-k') + i\lambda(k-k')
\end{matrix}
\right],
\end{split}
\en
\end{widetext}
where $k=(i\omega_n,\bk)$, $i\omega_n=i\pi T(2n+1)$ are the fermionic Matsubara frequencies. 
Note that the constraint field $i\lambda(k)$ plays the same role for the $f$-electron part of the propagator as a plasmon field $i\phi(k)$ for the $c$-electron part of the propagator. 
\begin{figure}[t]
\includegraphics[width=2.95in]{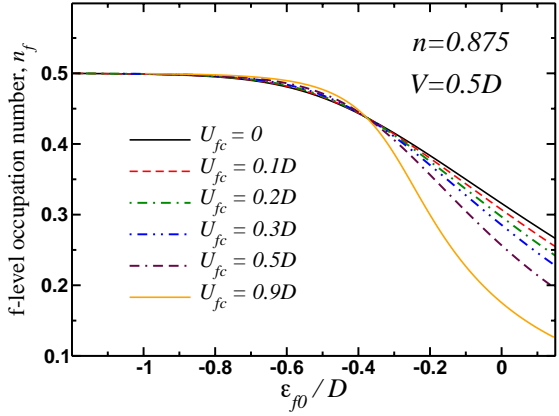}
\caption{Dependence of the $f$-level occupation number (per spin) on the position of the 'bare' $f$-level energy $\varepsilon_{f0}$ computed using the saddle-point approximation. All energies are given in the units of the band width $D$ of the conduction band. The values of the parameters used to obtain this plot are: $T=10^{-4}D$, $N=2$, $V=0.5D$, $n=0.875$. }
\label{Fig1nf}
\end{figure}

\subsection{Saddle-point approximation}
In the saddle-point approximation the bosonic fields are chosen in the following form:
\beg\label{SaddlePoint}
\begin{split}
&\rho(\bk,\tau)=\overline{\rho}\delta_{\bk,0}, \quad \lambda(\bk,\tau)=\overline{\lambda}\delta_{\bk,0}, \\
&\varphi_{\textrm{c,f}}(\bk,\tau)=\overline{\varphi}_{\textrm{c,f}}\delta_{\bk,0}, \quad \phi(\bk,\tau)=0.
\end{split}
\en
The zero value of the plasmon field $\phi(\bk,\tau)$ in (\ref{SaddlePoint}) means that its effect at the saddle-point level has already been included into the definition of the chemical potential. The stationary point equations in the saddle-point approximation are found by minimizing the action with respect to the bosonic fields (\ref{SaddlePoint}), which results in the system of the following equations:\cite{Miyake2000}
\beg\label{SaddlePointEqs}
\begin{split}
{i\overline{\lambda}}&=-T\sum\limits_{i\omega_n}\int_\bk\frac{V^2}{(i\omega_n-\overline{\varepsilon}_\bk)(i\omega_n-{\varepsilon}_f)-(V\overline{\rho})^2}, \\
q-\overline{\rho}^2&=T\sum\limits_{i\omega_n}\int_\bk\frac{(i\omega_n-\overline{\varepsilon}_\bk)e^{i\omega_n0+}}{(i\omega_n-\overline{\varepsilon}_\bk)(i\omega_n-{\varepsilon}_f)-(V\overline{\rho})^2}, \\
\overline{\varphi}_{\textrm{f}}&=2T\sum\limits_{i\omega_n}\int_\bk\frac{(i\omega_n-\overline{\varepsilon}_\bk)e^{i\omega_n0+}}{(i\omega_n-\overline{\varepsilon}_\bk)(i\omega_n-{\varepsilon}_f)-(V\overline{\rho})^2}, \\
\overline{\varphi}_{\textrm{c}}&=2T\sum\limits_{i\omega_n}\int_\bk\frac{(i\omega_n-{\varepsilon}_f)e^{i\omega_n0+}}{(i\omega_n-\overline{\varepsilon}_\bk)(i\omega_n-{\varepsilon}_f)-(V\overline{\rho})^2}.
\end{split}
\en
\begin{figure}[t]
\includegraphics[width=2.95in]{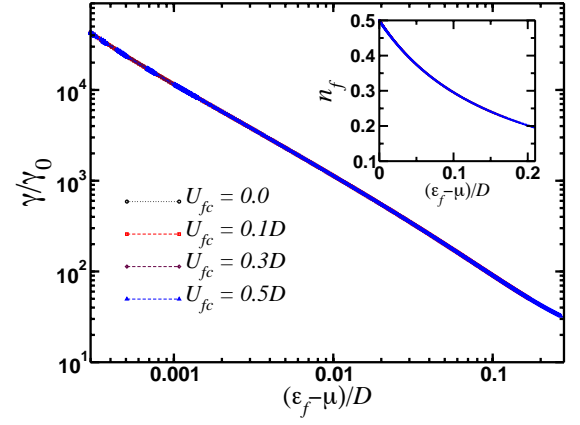}
\caption{Plot of the Sommerfeld coefficient $\gamma=(\pi^2/3)D\nu_F$ in the units of the Sommerfeld coefficient $\gamma_0$ in the absence of hybridization ($\nu_F$ is the single-particle density of states at the Fermi level) as a function of the Kondo lattice coherence temperature $T_{\textrm{coh}}=\varepsilon_f-\mu$ for various values of the coupling $U_{\textrm{fc}}$. The inset shows the dependence of the $f$-level occupation numbers as a function of $T_{\textrm{coh}}$. These results are found from the solution of the saddle-point equations assuming $T=10^{-4}D$, $N=2$, $V=0.5D$ and the total particle number $n=0.875$.}
\label{Fig2CT}
\end{figure}
In these equations $\int_\bk=\int\frac{d^3\bk}{(2\pi)^3}$, $\overline{\varepsilon}_\bk=\epsilon_\bk+U_{\textrm{fc}}\overline{\varphi}_{\textrm{f}}/2$ and ${\varepsilon}_f=\varepsilon_{f0}+U_{\textrm{fc}}\overline{\varphi}_{\textrm{c}}/2+i\overline{\lambda}$.
From the last two equations (\ref{SaddlePointEqs}) it follows that $\overline{\varphi}_{\textrm{c,f}}=2{n}_{\textrm{c,f}}$, where ${n}_\alpha$ are the average occupation numbers per spin. 
In what follows we assume that the total number of particles is fixed
\beg\label{totalpn}
\overline{n}={n}_c+{n}_f=\textrm{const.}
\en
The Matsubara summations can be easily performed using the Poisson summation formula
\beg\label{Poisson}
T\sum\limits_{i\omega_n}\frac{1}{(i\omega_n+i\nu-a)(i\omega_n-b)}=\frac{n_F(b)-n_F(a)}{i\nu+b-a},
\en
where $n_F(\varepsilon)=1/\left(\exp[(\varepsilon-\mu)/T]+1\right)$ is the Fermi distribution function and $\mu$ is a chemical potential. As a result, we obtain the following three equations which self-consistently determine the values of ${\varepsilon}_f$, $\overline{\rho}$ and the chemical potential $\mu$:
\beg\label{MainMFEqs}
\begin{split}
&{\varepsilon}_f-\varepsilon_{f0}-U_{\textrm{fc}}n_{c}=\int_\bk\frac{n_F(E_{2\bk})-n_F(E_{1\bk})}{\sqrt{(\overline{\varepsilon}_f-\overline{\varepsilon}_\bk)^2+(2V\overline{\rho})^2}}, \\
&n_f+n_c=\int_\bk\left[n_F(E_{1\bk})+n_F(E_{2\bk})\right], \\
&n_f-n_c=\int_\bk\frac{(\overline{\varepsilon}_\bk-{\varepsilon}_f)[n_F(E_{2\bk})-n_F(E_{1\bk})]}{\sqrt{({\varepsilon}_f-\overline{\varepsilon}_\bk)^2+(2V\overline{\rho})^2}},
\end{split}
\en
where we introduced
\beg\label{E12k}
E_{1(2)\bk}=\frac{1}{2}\left(\overline{\varepsilon}_\bk+{\varepsilon}_f\pm\sqrt{({\varepsilon}_f-\overline{\varepsilon}_\bk)^2+4(V\overline{\rho})^2}\right).
\en
\begin{figure}[t]
\includegraphics[width=2.5in]{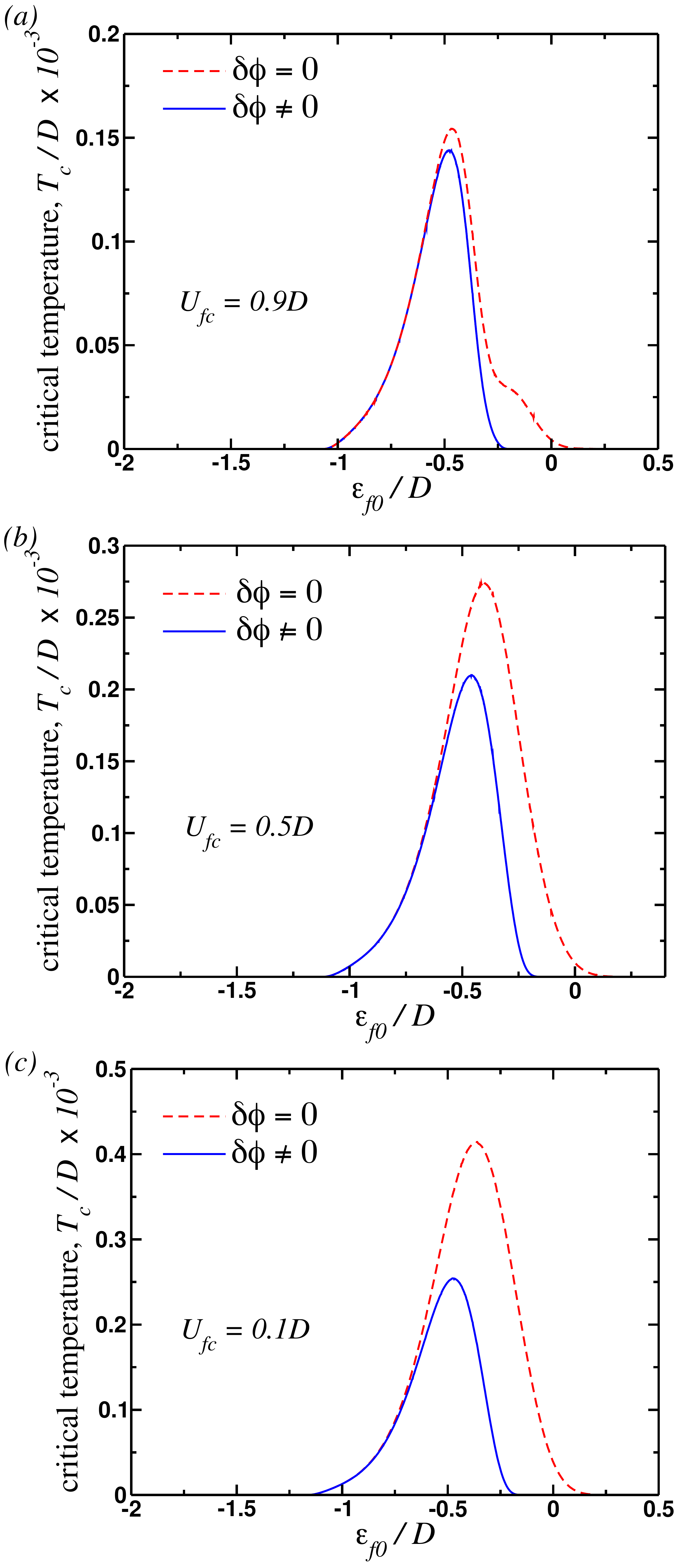}
\caption{Superconducting critical temperature computed for three different values of the Hubbard $U_{\textrm{fc}}$ interaction and for  the two separate cases of zero and non-zero plasmonic field fluctuations. We find that the fluctuations of the plasmonic field has significant effect on critical temperature when $U_{\textrm{fc}}\ll D$. Contrary to the earlier studies, we found that the maximum value of the critical temperature is decreasing with the increase in the strength of $U_{\textrm{fc}}$. These results are found from the solution of the saddle-point equations with $p_F=(2mE_F/D)^{1/2}\approx3.73$, $T=10^{-5}D$, $N=2$, $V=0.5D$ and the total particle number $n=0.875$.}
\label{Fig3}
\end{figure}

To perform the integration over momentum, we will assume that the particle density in the conduction band is low enough. Thus, we consider the Galilean-invariant spectrum 
\beg\label{ek}
\eps_\bk=\frac{\bk^2}{2m}-D,
\en
where $D$ is set as a unit of energy. The effective mass $m=(3\pi^2/\sqrt{2})^{2/3}/2D$ of the conduction electrons is obtained from the condition that the total number of particles (per spin) equals one:
\beg\label{IntPart}
\int\limits_{-D}^{D}\nu_0(\eps_\bk)d\eps_\bk=1.
\en
In this formula $\nu_0(\eps)=(3/4\sqrt{2}D)\sqrt{\eps/D+1}$ is the single-particle density of states for the non-interacting system. In what follows we will limit out calculations to the case when the particle occupation number of the conduction band (per spin) equals to $0.375$, which means that the total particle occupation number per spin must be $n=0.875$. 

The results of the numerical solution of these equations are presented in Fig. 1 and Fig. 2. In Fig. 1 we show the dependence of the $f$-level occupation numbers as a function of the bare $f$-level energy. Notice, that with an increase in the values of $U_{\textrm{fc}}$ the system approaches the first order phase transition when the value of $n_f$ changes abruptly from $n_f^{(\textrm{interm.})}\sim 0.1$ to $n_f^{(\textrm{local})}\sim 0.5$. In Fig. 2 we plot the dependence of the Sommerfeld coefficient $\gamma=\lim\limits_{T\to 0}C(T)/T$ as a function of the parameter
$\varepsilon_f-\mu$ (the latter is usually associated with the coherence temperature of the Kondo lattice, $T_{\textrm{coh}}$). The value of $\gamma$ is significantly enhanced in the local moment regime $n_f\sim 0.5$. It is worth pointing out that at least in the local moment regime the fluctuations associated with the plasmon field $\phi(\bk,\tau)$ will not affect the value of $\gamma$ significantly, since their contribution is proportional to $1/N$. Since $T_{\textrm{coh}}\sim\nu_F^{-1}$ and both of these parameters will ultimately determine the value of the superconducting transition temperature, it is clear that the effects associated with the plasmon field will be encoded into the magnitude of the pairing strength. Furthermore, the Fermi energy is given by 
\beg\label{NewFermi}
E_F=\mu+\frac{(V\overline{\rho})^2}{T_{\textrm{coh}}}-U_{\textrm{fc}}n_f
\en
and so the Fermi energy increases with an increase in effective hybridization, $V\overline{\rho}$.
Overall, our results presented in Figs 1 and Fig. 2 agree with those reported earlier.\cite{Miyake2000}
\subsection{Fluctuation propagator}
Having determined the values of the bosonic fields at the stationary point, we can determine the propagators of the bosonic fields at the gaussian level. We represent the bosonic fields as
\beg\label{Expand}
\begin{split}
&\rho(\bk,\tau)=\overline{\rho}\delta_{\bk,0}+\delta\rho(\bk,\tau),  ~i\phi(\bk,\tau)=\delta(i\phi(\bk,\tau)), \\
&i\lambda(\bk,\tau)=i\overline{\lambda}\delta_{\bk,0}+\delta(i\lambda(\bk,\tau)), \\
&\varphi_{\textrm{c,f}}(\bk,\tau)=\overline{\varphi}_{\textrm{c,f}}\delta_{\bk,0}+\delta\varphi_{\textrm{c,f}}(\bk,\tau), \\
\end{split}
\en 
and expand each term in the action (\ref{BosonicAction}) in powers of the components of 
\beg\label{BigPhi}
\delta\hat{\Phi}=\left(\delta\rho, \delta(i\lambda), \delta\varphi_{\textrm{c}}, \delta\varphi_{\textrm{f}},\delta(i\phi)\right)^t
\en
up to the second order ($t$ means transpose). For the details on the derivation we refer the reader to Appendix A. As a result, an inverse of the fluctuation propagator can be represented in terms of a $5\times 5$ matrix given by 
\begin{widetext}
\beg\label{Sq}
\hat{\cal S}_q=\left[
\begin{matrix}
i\overline{\lambda}+V^2\Pi_{\textrm{vv}}(q)+V^2\Pi_{3}(q) & \overline{\rho}+V\Pi_2(q) & 
\left(\frac{VU_{\textrm{fc}}}{2}\right)\Pi_2(q) & \left(\frac{VU_{\textrm{fc}}}{2}\right)\Pi_1(q) & V\Pi_1(q) \\
\overline{\rho}+V\Pi_2(-q) &  \frac{1}{2}\Pi_{\textrm{ff}}(q) & \left(\frac{U_{\textrm{fc}}}{4}\right)\Pi_{\textrm{ff}}(q) & 
\left(\frac{U_{\textrm{fc}}}{4}\right)\Pi_{\textrm{vv}}(q) & \frac{1}{2}\Pi_{\textrm{vv}}(q) \\
\left(\frac{VU_{\textrm{fc}}}{2}\right)\Pi_2(-q) & \left(\frac{U_{\textrm{fc}}}{4}\right)\Pi_{\textrm{ff}}(-q) & \left(\frac{U_{\textrm{fc}}^2}{8}\right)\Pi_{\textrm{ff}}(q) & 
-\frac{U_{\textrm{fc}}}{8}+\left(\frac{U_{\textrm{fc}}^2}{8}\right)\Pi_{\textrm{vv}}(q) & \left(\frac{U_{\textrm{fc}}}{4}\right)\Pi_{\textrm{vv}}(q) \\
\left(\frac{VU_{\textrm{fc}}}{2}\right)\Pi_1(-q) & \left(\frac{U_{\textrm{fc}}}{4}\right)\Pi_{\textrm{vv}}(-q) & -\frac{U_{\textrm{fc}}}{8}+\left(\frac{U_{\textrm{fc}}^2}{8}\right)\Pi_{\textrm{vv}}(-q) & \left(\frac{U_{\textrm{fc}}^2}{8}\right)\Pi_{\textrm{cc}}(q) & 
\left(\frac{U_{\textrm{fc}}}{4}\right)\Pi_{\textrm{cc}}(q) \\ \left(\frac{VU_{\textrm{fc}}}{2}\right)\Pi_1(-q) & \frac{1}{2}\Pi_{\textrm{vv}}(-q) & \left(\frac{U_{\textrm{fc}}}{4}\right)\Pi_{\textrm{vv}}(-q) & \left(\frac{U_{\textrm{fc}}}{4}\right)\Pi_{\textrm{cc}}(-q) & -\frac{1}{2U(\bq)}+\frac{1}{2}\Pi_{\textrm{cc}}(q)
\end{matrix}
\right].
\en
\end{widetext}
The corresponding expressions for the polarization operators entering into this expression can be found in the Appendix B. It is worth noting here that not all of the polarization operators are independent. For example, a simple calculation shows that
\beg\label{Pi3simple}
V^2\Pi_3(q)=-i\overline{\lambda}+V^2\Pi_{\textrm{vv}}(q)-i\nu\frac{V}{\overline{\rho}}\Pi_2(q),
\en
where $q=(\bq,i\nu)$. Finally, a quantity which will be central to our discussion below - bosonic propagator - is given by the inverse of (\ref{Sq}):
\beg\label{Dq}
\hat{\cal D}(q)=-\hat{\cal S}_q^{-1}.
\en
In what follows we will use equations (\ref{Sq},\ref{Dq}) to investigate the superconducting instability mediated by the interactions between the fermions and fluctuating bosonic fields. 

\section{Superconductivity from repulsive electron-electron interactions}
The problem of superconductivity emerging as a ground state in the Anderson lattice model has been extensively discussed in the literature starting with the pioneering papers by Lavagna, Millis and Lee [\onlinecite{Lavagna1987}] and by Houghton, Read and Won [\onlinecite{Read1988}]. In the context of the extended Anderson model the problem of superconducting pairing mediated by the bosonic fluctuations has been studied by Onishi and Miyake [\onlinecite{Miyake2000}]. Specifically, they found that with an increase in the strength of the Hubbard interaction $U_{\textrm{fc}}$ between the conduction and $f$-electrons, the critical temperature of the superconducting transition also increases, i.e. increasing the strength of the local repulsive interaction boosts superconductivity. Since there are no retardation effects and all interactions are repulsive, it is expected that the  superconducting order parameter has nodes and the highest transition temperature was found to be for the $d$-wave symmetry. In this regard, it will be interesting to check whether the long-range Coulomb interactions may produce the same effect here. 

By the nature of the interaction which induces the Cooper pairing, in the weak coupling theory the Kondo lattice coherence temperature $T_{\textrm{coh}}$ plays a role of the characteristic energy scale analogous to the Debye frequency in the conventional theory of superconductivity. The critical temperature describing the superconducting instability in the $l$-orbital channel is given by 
\beg\label{Tc}
T_c^{(l)}=T_{\textrm{coh}}e^{-1/\lambda_l}, 
\en
where $\lambda_l=\nu_F\Gamma_l$ is the dimensionless coupling constant, $\nu_F$ is the density of states at the Fermi level and $\Gamma_l>0$ is given by
\beg\label{Gammal}
\Gamma_l=\left(\frac{2l+1}{2}\right)\int\limits_0^\pi\Gamma^{(0)}(\theta)P_l(\cos\theta)\sin\theta d\theta.
\en
Here $P_l(\cos\theta)$ is the Legendre polynomial and $\Gamma^{(0)}(\theta)$ the bare interaction in the Cooper channel.\cite{AGD} 

Interaction function $\Gamma^{(0)}(\theta)$ is determined by the matrix elements of the fluctuation propagator $\hat{\cal D}(\bq,i\nu)$ evaluated at $|\bq|=2k_F\sin(\theta/2)$ and $i\nu\to 0$.\cite{Lavagna1987,Read1988,Miyake2000} The specific form of 
$\Gamma^{(0)}(\theta)$ depends on whether the chemical potential is in the first $E_{1\bk}$ or the second $E_{2\bk}$ band. Since we have chosen the fairly low occupation number for the conduction band, we find that the chemical potential lies close to the top of the second band, $E_{2\bk}$. 
The fermionic operators $a_{\bk\sigma}, a_{\bk\sigma}\dg$, which describe the quasiparticles in this band are related to the original fermionic operators $c_{\bk\sigma}$ and $f_{\bk\sigma}$ by the following relation:
\beg\label{gamma}
\begin{split}
c_{\bk\sigma}=-v_\bk a_{\bk\sigma}, \quad f_{\bk\sigma}=u_\bk a_{\bk\sigma}.
\end{split}
\en
Here $u_\bk$ and $v_\bk$ are the coherence factors defined in the Appendix B. We introduce the two-particle correlation function 
\beg\label{TwoParticle}
\Gamma_{\alpha\beta\gamma\delta}(12;34)=-\left\langle\hat{T}_\tau\left\{a_{\alpha}(1)a_{\beta}(2){a}_{\gamma}\dg(3)\overline{a}_{\delta}\dg(4)\right\}\right\rangle,
\en
where averaging is performed over the action (\ref{BosonicAction}) and $a_\sigma(1)=a_\sigma(\bk_1,\tau_1)$ etc. Expanding the action up to the second order in $\delta\Phi$ (\ref{BigPhi}), we have found the following expression for the interaction function:
\begin{widetext}
\beg\label{Interaction}
\begin{split}
&\Gamma^{(0)}(\theta)=v_{k_F}^4\left[\left(\frac{U_{\textrm{fc}}}{2}\right)^2{\cal D}_{\varphi_{\textrm{f}}\varphi_{\textrm{f}}}(\theta)+
\frac{U_{\textrm{fc}}}{2}{\cal D}_{\varphi_{\textrm{f}}\phi}(\theta)+\frac{U_{\textrm{fc}}}{2}{\cal D}_{\phi\varphi_{\textrm{f}}}(\theta)+
{\cal D}_{\phi\phi}(\theta)\right]+u_{k_F}^4\left[{\cal D}_{\lambda\lambda}(\theta)+\frac{U_{\textrm{fc}}}{2}{\cal D}_{\lambda\varphi_{\textrm{c}}}(\theta)\right.\\&\left.+\frac{U_{\textrm{fc}}}{2}{\cal D}_{\varphi_{\textrm{c}}\lambda}(\theta)+
\left(\frac{U_{\textrm{fc}}}{2}\right)^2{\cal D}_{\varphi_{\textrm{c}}\varphi_{\textrm{c}}}(\theta)\right]+2u_{k_F}^2v_{k_F}^2\left[
\frac{U_{\textrm{fc}}}{2}{\cal D}_{\lambda\varphi_{\textrm{f}}}(\theta)+\left(\frac{U_{\textrm{fc}}}{2}\right)^2{\cal D}_{\varphi_{\textrm{c}}\varphi_{\textrm{f}}}(\theta)+\frac{U_{\textrm{fc}}}{2}{\cal D}_{\varphi_{\textrm{c}}\phi}(\theta)+{\cal D}_{\lambda\phi}(\theta)\right.\\&\left.+2V^2{\cal D}_{\rho\rho}(\theta)\right]-4u_{k_F}v_{k_F}^3\left[V{\cal D}_{\phi\rho}(\theta)+\frac{VU_{\textrm{fc}}}{2}{\cal D}_{\varphi_{\textrm{f}}\rho}(\theta)\right]-4v_{k_F}u_{k_F}^3\left[V{\cal D}_{\lambda\rho}(\theta)+\frac{VU_{\textrm{fc}}}{2}{\cal D}_{\varphi_{\textrm{c}}\rho}(\theta)\right]\\&\equiv \Gamma_{\textrm{cccc}}(\theta)+\Gamma_{\textrm{ffff}}(\theta)+
\Gamma_{\textrm{ffcc}}(\theta)+\Gamma_{\textrm{cccf}}(\theta)+\Gamma_{\textrm{fffc}}(\theta)
\end{split}
\en
\end{widetext}
and the coherence factors $u_{k_F}$, $v_{k_F}$ are evaluated at the Fermi energy: 
\beg\label{ukFvkF}
u_{k_F}=\frac{V\overline{\rho}}{\sqrt{T_{\textrm{coh}}^2+(V\overline{\rho})^2}}, 
~v_{k_F}=\frac{T_{\textrm{coh}}}{\sqrt{T_{\textrm{coh}}^2+(V\overline{\rho})^2}}.
\en
The subscripts in ${\cal D}$ refer to its matrix elements, i.e.  ${\cal D}_{\varphi_{\textrm{c}}\lambda}$ corresponds to the matrix element ${\cal D}_{32}$ etc. in accordance with the definition (\ref{BigPhi}).

The relations (\ref{ukFvkF}) imply that in the vicinity of the local moment regime $n_f\sim 1/2$, when $T_{\textrm{coh}}\ll D$ we find that $u_{k_F}\sim1$.
Therefore, we can expect that the largest contribution to the pairing interaction is provided by the last term in (\ref{Interaction}), which is confirmed by the numerical calculations. This means that the long-range Coulomb interactions should not significantly affect the value of the transition temperature in the local moment regime: in comparison to the contribution from the itinerant $c$-states, the $f$-states provide significantly larger spectral weight contribution to the pairing quasiparticles. 

Our results of the numerical calculation of the superconducting critical temperature are shown in Fig. 3. First, we found that the largest critical temperature is realized in the $d$-wave ($l=2$) channel. In agreement with the previous studies, we have found that the maximum of the critical temperature remains weakly dependent on the value of $\veps_{f0}$. At the same time we find that the critical temperature is decreasing with both an inclusion of the plasmon field fluctuations as well as with an increase in the values of $U_{\textrm{fc}}$, which is strike contrast with the earlier results.\cite{Miyake2000}

\begin{figure}[t]
\includegraphics[width=2.5in]{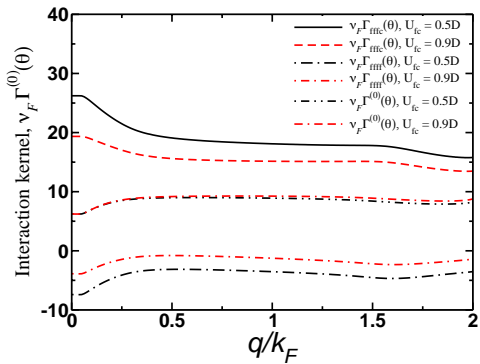}
\caption{Interaction kernel as a function of momentum computed without the effects of the plasmon fluctuations. This result shows that although separate contributions to $\Gamma^{(0)}$, such as the leading ones $\Gamma_{\textrm{fffc}}$ and $\Gamma_{\textrm{ffff}}$, do depend on the value of $U_{\textrm{fc}}$, $\Gamma^{(0)}$ shows much weaker dependence on $U_{\textrm{fc}}$. Notably, as the value of $U_{\textrm{fc}}$ increases, the effective interaction becomes more repulsive rendering the lower $T_c$ in the $d$-wave channel.  These results are found from the solution of the saddle-point equations with $k_F=(2mE_F/D)^{1/2}\approx3.73$, $T=10^{-5}D$, $N=2$, $V=0.5D$ and the total particle number $n=0.875$. The values of the remaining parameters are as follows:
for $U_{\textrm{fc}}=0.5D$: $\veps_{f0}=-0.40487D$, 
$\veps_f=0.3461D$,  $\overline{\rho}=0.2304$ and $\mu=0.3279D$ and for $U_{\textrm{fc}}=0.9D$ and $\veps_{f0}=-0.46915D$
we found $\veps_f=0.50685D$,  $\overline{\rho}=0.17148$ and $\mu= 0.49712D$.}
\label{Fig4}
\end{figure}

In order to get insight into the origin of this effect, we fix the values of $\veps_f$, $\overline{\rho}$ and $\mu$ to their corresponding values at the maximum of $T_c$ and consider how the interaction (\ref{Interaction}) changes with the changes in $U_{\textrm{fc}}$. 
First of all, we recall that $\nu_F$ remains independent on the value of $U_{\textrm{fc}}$, Fig. \ref{Fig2CT}. This result implies that $T_{\textrm{coh}}$ must also remain essentially independent of $U_{\textrm{fc}}$. Indeed, for the results shown in Fig. \ref{Fig3} we found that $T_{\textrm{coh}}/D\approx 2.3\times 10^{-3}$ and $\nu_F/\nu_0\approx 770$ when $T_c$ approaches its maximum value.  At the same time two largest contributions to the interaction kernel, $\Gamma_{\textrm{fffc}}(\theta)$ and $\Gamma_{\textrm{ffff}}(\theta)$, do change with $U_{\textrm{fc}}$, Fig. \ref{Fig4}. As it turns, however, their respective contributions to the $\nu_F\Gamma^{(0)}(\theta)$ yield similar results and, as a consequence, we find that $\nu_F\Gamma^{(0)}(\theta)$ is slightly increased while the value of the maximum $T_c$ is somewhat reduced with an increase of $U_{\textrm{fc}}$. In other words, valence fluctuations produce superconductivity even in the case of when $U_{\textrm{fc}}$ can be neglected and the increasing the value of $U_{\textrm{fc}}$ reduced the value of the superconducting critical temperature to an intermediate valence regime. The same argument can be implied to the effects of the long-range Coulomb interactions, which also lead to the reduction in the value of $T_c$ as soon as the system in tuned away from the local moment regime. 

\section{Conclusions}
In this paper we have presented the results of the calculations for the critical temperature of the superconducting transition in the extended Anderson lattice model, which also included the long-range Coulomb repulsion between the conduction electrons. Our work has been motivated by recent discovery of the Ce-based cage compounds CeNi$_2$Cd$_{20}$ and CePd$_2$Cd$_{20}$. These compounds have vanishing RKKY interactions and do not exhibit a long-range order down to very low temperatures in the millikelvin range. We propose that $d$-wave superconductivity may develop in these compounds under an application of the hydrostatic pressure. 

We have also found that the long-range Coulomb repulsion leads to a decrease in the values of superconducting critical temperature: in the local moment regime it has small effect on the value of $T_c$ due to the fact the most of the spectral weight carried by the quasiparticles, which form the Cooper pairs, is provided by the $f$-electrons. In the mixed-valent regime and for the low enough values of $U_{\textrm{fc}}$ the fluctuations of the plasmon field significantly reduces the value of the superconducting critical temperature. 

\section{Acknowledgments} The authors would like to thank C. C. Almasan and M. B. Maple for bringing their attention to this problem. We acknowledge useful discussions with B. Fregoso and Y. Li related to this project. This work was financially supported by the National Science Foundation grant NSF-DMR-2002795.
\begin{appendix}
\section{Effective action in the gaussian approximation}
We define the single-particle fermionic propagator in the stationary point, which is obtained from (\ref{PropG}) using (\ref{SaddlePoint})
\beg\label{PropGsp}
\hat{\cal G}^{-1}(k)=\left[
\begin{matrix}
i\omega_n-\eps_\bk-U_{\textrm{fc}}n_f & -V\overline{\rho} \\ -V\overline{\rho} & 
i\omega_n-\veps_{f}
\end{matrix}
\right].
\en
Then the action at the stationary point is given by 
\beg\label{S0}
\begin{split}
S_0&=-N\textrm{Tr}\log\left[-\hat{\cal G}^{-1}\right]+iN\overline{\lambda}\left(\overline{\rho}^2-q_0\right)\\&-\frac{NU_{\textrm{fc}}}{4}\overline{\varphi}_{\textrm{f}}\overline{\varphi}_{\textrm{c}}.
\end{split}
\en
Minimizing $S_0$ with respect to the slave-boson fields yields the equations (\ref{SaddlePointEqs}).

We can now expand action (\ref{BosonicAction}) up to the second order in powers of $\delta\Phi$ (see (\ref{Expand}) in main text).
It follows
\beg\label{SS0}
S=S_0+\delta S,
\en
where the fluctuation correction to the action $\delta S$ is of the form
\begin{widetext}
\beg\label{deltaS}
\begin{split}
\delta S&=-N\textrm{Tr}\log\left[1-\hat{\cal G}\hat{\cal M}\right]
+N\sum\limits_k\left[(i\overline{\lambda})\delta\rho(-k)\delta\rho(k)+2\overline{\rho}i\delta\lambda(-k)\delta \rho(k)\right]\\-
&N\sum\limits_k\left[\frac{U_{\textrm{fc}}}{4}\delta\varphi_{\textrm{f}}(-k)\delta\varphi_{\textrm{c}}(k)+\frac{\delta(i\phi(-k))\delta(i\phi(k))}{U(\bk)}\right].
\end{split}
\en
Here matrix $\hat{\cal M}$ is defined by
\beg\label{MatM}
\hat{\cal M}_{kk'}=\left[\begin{matrix}
\frac{U_{\textrm{fc}}}{2}\delta\varphi_{\textrm{f}}(k-k')+\delta(i\phi(k-k')) & V\delta\rho(k-k') \\ V\delta\rho(k-k') & 
\frac{U_{\textrm{fc}}}{2}\delta\varphi_{\textrm{c}}(k-k')+\delta(i\lambda(k-k'))
\end{matrix}\right].
\en
\end{widetext}
In the gaussian approximation we formally expand the expression under the logarithm (\ref{deltaS})
\beg\label{dSExpand}
-\textrm{Tr}\log\left[1-\hat{\cal G}\hat{\cal M}\right]=\sum\limits_{n=1}^{\infty}\frac{1}{n}\textrm{Tr}\left(\hat{\cal G}\hat{\cal M}\right)^n
\en
and retain only the term with $n=2$. As a result we find the following expression for (\ref{deltaS}):
\beg\label{FindeltaS}
\delta S=N\sum\limits_k\delta\hat{\Phi}(-k)\hat{\cal S}_{k}\hat{\Phi}(k),
\en
with $\hat{\cal S}_k$ given by Eq. (\ref{Sq}) in the main text. Upon integrating out the bosonic fields, it is straightforward to check that the fluctuations give the correction of the order of $O(1/N)$ to the free energy. 
\section{Polarization operators}
The polarization operators which enter into equation (\ref{Sq}) are defined according to
\beg\label{PolOps}
\begin{split}
&\Pi_{\textrm{cc}}(q)=T\sum\limits_{i\omega_n}\sum\limits_\bk{\cal G}_{\textrm{cc}}(k+q){\cal G}_{\textrm{cc}}(k), \\
&\Pi_{\textrm{ff}}(q)=T\sum\limits_{i\omega_n}\sum\limits_\bk{\cal G}_{\textrm{ff}}(k+q){\cal G}_{\textrm{ff}}(k), \\
&\Pi_{\textrm{vv}}(q)=T\sum\limits_{i\omega_n}\sum\limits_\bk{\cal G}_{\textrm{fc}}(k+q){\cal G}_{\textrm{cf}}(k),\\
&\Pi_{\textrm{cv}}(q)=T\sum\limits_{i\omega_n}\sum\limits_\bk{\cal G}_{\textrm{cc}}(k+q){\cal G}_{\textrm{cf}}(k)=\Pi_{\textrm{vc}}(-q), \\
&\Pi_{\textrm{fv}}(q)=T\sum\limits_{i\omega_n}\sum\limits_\bk{\cal G}_{\textrm{ff}}(k+q){\cal G}_{\textrm{fc}}(k)
=\Pi_{\textrm{vf}}(-q), \\
&\Pi_{\textrm{fc}}(q)=T\sum\limits_{i\omega_n}\sum\limits_\bk{\cal G}_{\textrm{ff}}(k+q){\cal G}_{\textrm{cc}}(k)
=\Pi_{\textrm{cf}}(-q).
\end{split}
\en
Here $q=(\bq,i\nu_m)$, $i\nu_m=2i\pi Tm$ is the bosonic Matsubara frequency, $k=(\bk,i\omega_n)$ and $i\omega_n=i\pi T(2n+1)$
is the fermionic Matsubara frequency. Functions ${\cal G}_{\textrm{aa}}(k)$ which appear in this expressions are defined according to
\beg\label{DefGab}
\begin{split}
{\cal G}_{\textrm{cc}}(k+q)&=\frac{i\omega_n-\varepsilon_f}{(i\omega_n-\overline{\varepsilon}_\bk)(i\omega_n-{\varepsilon}_f)-(V\overline{\rho})^2}\\&=\frac{u_\bk^2}{i\omega_n-E_{1\bk}}+\frac{v_\bk^2}{i\omega_n-E_{2\bk}}, \\
{\cal G}_{\textrm{ff}}(k+q)&=\frac{i\omega_n-\overline{\varepsilon}_\bk}{(i\omega_n-\overline{\varepsilon}_\bk)(i\omega_n-{\varepsilon}_f)-(V\overline{\rho})^2}\\&=\frac{v_\bk^2}{i\omega_n-E_{1\bk}}+\frac{u_\bk^2}{i\omega_n-E_{2\bk}}, \\
\end{split}
\en
The remaining correlator ${\cal G}_{\textrm{fc}}={\cal G}_{\textrm{cf}}$ is given by 
\beg\label{Gv}
\begin{split}
&{\cal G}_{\textrm{fc}}(k+q)=\frac{V^2}{(i\omega_n-\overline{\varepsilon}_\bk)(i\omega_n-{\varepsilon}_f)-(V\overline{\rho})^2}\\&=\frac{V\overline{\rho}}{E_{1\bk}-E_{2\bk}}\left(\frac{1}{i\omega_n-E_{1\bk}}-\frac{1}{i\omega_n-E_{2\bk}}\right).
\end{split}
\en
In the expressions above we introduced the coherence factors
\beg\label{uk2vk2}
\begin{split}
u_\bk^2=\frac{1}{2}\left(1+\frac{\overline{\veps}_\bk-\veps_f}{R_\bk}\right), 
~v_\bk^2=\frac{1}{2}\left(1-\frac{\overline{\veps}_\bk-\veps_f}{R_\bk}\right)
\end{split}
\en
and $R_\bk=E_{1\bk}-E_{2\bk}$.
For convenience, instead of the last three polarization functions, we will consider 
\beg\label{Pi123}
\begin{split}
\Pi_1(\bq,i\nu_l)&=\frac{1}{2}\left[\Pi_{\textrm{cv}}(\bq,i\nu_l)+\Pi_{\textrm{vc}}(\bq,i\nu_l)\right], \\
\Pi_2(\bq,i\nu_l)&=\frac{1}{2}\left[\Pi_{\textrm{fv}}(\bq,i\nu_l)+\Pi_{\textrm{vf}}(\bq,i\nu_l)\right], \\
\Pi_3(\bq,i\nu_l)&=\frac{1}{2}\left[\Pi_{\textrm{cf}}(\bq,i\nu_l)+\Pi_{\textrm{fc}}(\bq,i\nu_l)\right].
\end{split}
\en
The summations over the Matsubara frequencies can be easily performed using (\ref{Poisson}). For example
\begin{widetext}
\beg\label{SumPcc}
\begin{split}
&T\sum\limits_{i\omega_n}\frac{(i\omega_{n+l}-\veps_{\bk+\bq})}{(i\omega_{n+l}-E_{1\bk+\bq}+\mu)
(i\omega_{n+l}-E_{2\bk+\bq}+\mu)}\frac{(i\omega_{n}-\veps_f)}{(i\omega_n-E_{1\bk}+\mu)(i\omega_n-E_{2\bk}+\mu)}\\&=
v_{\bk+\bq}^2u_\bk^2\frac{[n_F(E_{1\bk})-n_F(E_{1{\bk+\bq}})]}{i\nu_l+E_{1\bk}-E_{1\bk+\bq}}+
v_{\bk+\bq}^2v_\bk^2\frac{[n_F(E_{2\bk})-n_F(E_{1{\bk+\bq}})]}{i\nu_l+E_{2\bk}-E_{1\bk+\bq}}
+u_{\bk+\bq}^2u_\bk^2\frac{[n_F(E_{1\bk})-n_F(E_{2{\bk+\bq}})]}{i\nu_l+E_{1\bk}-E_{2\bk+\bq}}\\&+
u_{\bk+\bq}^2v_\bk^2\frac{[n_F(E_{2\bk})-n_F(E_{2{\bk+\bq}})]}{i\nu_l+E_{2\bk}-E_{2\bk+\bq}}.
\end{split}
\en
\end{widetext}
We also would like to remind the reader that all the energies entering into these expressions are taken relative to the chemical potential $\mu$. Formally, this is accomplished by including the chemical potential into the definition of the Fermi distribution function, 
Eq. (\ref{Poisson}).

\end{appendix}

%\bibliography{uccLNbib}

\begin{thebibliography}{100}

\bibitem{Burnett2014}
V.~Burnett, D.~Yazici, B.~White, N.~Dilley, A.~Friedman, B.~Brandom, and
  M.~Maple, ``Structure and physical properties of {RT}$_2$Cd$_{20}$ (R=rare earth,
  T=Ni, Pd) compounds with the {CeCr}$_2$al$_{20}$-type structure,'' {\em
  Journal of Solid State Chemistry}, vol.~215, pp.~114--121, jul 2014.

\bibitem{Yazici2015}
D.~Yazici, T.~Yanagisawa, B.~D. White, and M.~B. Maple, ``Nonmagnetic ground
  state in the cubic compounds {PrNi}$_2$Cd$_{20}$ and PrPd$_2$Cd$_{20}$,'' {\em Physical Review
  B}, vol.~91, p.~115136, mar 2015.

\bibitem{Onimaru2010}
T.~Onimaru, K.~T. Matsumoto, Y.~F. Inoue, K.~Umeo, Y.~Saiga, Y.~Matsushita,
  R.~Tamura, K.~Nishimoto, I.~Ishii, T.~Suzuki, and T.~Takabatake,
  ``Superconductivity and structural phase transitions in caged compounds
  {RT}$_2$Zn$_{20}$ (r = La, Pr, t = Ru, Ir),'' {\em Journal of the Physical
  Society of Japan}, vol.~79, p.~033704, Mar 2010.

\bibitem{Niemann1995}
S.~Niemann and W.~Jeitschko, ``Ternary aluminides {AT}$_2$Al$_{20}$ (A = rare earth
  elements and uranium: T = Ti, Nb, Ta, Mo, and W) with {CeCr}$_2$Al$_{20}$-type
  structure,'' {\em Journal of Solid State Chemistry}, vol.~114, pp.~337--341,
  Feb 1995.

\bibitem{Kangas2012}
M.~J. Kangas, D.~C. Schmitt, A.~Sakai, S.~Nakatsuji, and J.~Y. Chan,
  ``Structure and physical properties of single crystal {PrCr}$_2$Al$_{20}$ and
  {CeM}$_2$Al$_{20}$ (M=V, Cr): A comparison of compounds adopting the {CeCr}2al20
  structure type,'' {\em Journal of Solid State Chemistry}, vol.~196,
  pp.~274--281, Dec 2012.

\bibitem{Isikawa2013}
Y.~Isikawa, T.~Mizushima, K.~Kumagai, and T.~Kuwai, ``Dense Kondo effect in
  caged compound {CeRu}$_2$Zn$_{20}$,'' {\em Journal of the Physical Society of
  Japan}, vol.~82, p.~083711, Aug 2013.

\bibitem{Swatek2013}
P.~Swatek and D.~Kaczorowski, ``Intermediate valence behavior in the novel cage
  compound {CeIr}$_2$Zn$_{20}$,'' {\em Journal of Physics: Condensed Matter},
  vol.~25, p.~055602, Jan 2013.

\bibitem{White2015}
B.~D. White, D.~Yazici, P.-C. Ho, N.~Kanchanavatee, N.~Pouse, Y.~Fang, A.~J.
  Breindel, A.~J. Friedman, and M.~B. Maple, ``Weak hybridization and isolated
  localized magnetic moments in the compounds {CeT}$_2$Cd$_{20}$(T = Ni, Pd),'' {\em
  Journal of Physics: Condensed Matter}, vol.~27, p.~315602, Jul 2015.

\bibitem{AndersonSuper}
P.~W. Anderson, ``Antiferromagnetism. theory of superexchange interaction,''
  {\em Phys. Rev.}, vol.~79, pp.~350--356, Jul 1950.

\bibitem{Ruderman1954}
M.~A. Ruderman and C.~Kittel, ``Indirect exchange coupling of nuclear magnetic
  moments by conduction electrons,'' {\em Physical Review}, vol.~96,
  pp.~99--102, oct 1954.

\bibitem{Kasuya1956}
T.~Kasuya, ``A theory of metallic ferro- and antiferromagnetism on
  zener{\textquotesingle}s model,'' {\em Progress of Theoretical Physics},
  vol.~16, pp.~45--57, jul 1956.

\bibitem{Yosida1957}
K.~Yosida, ``Magnetic properties of Cu-Mn alloys,'' {\em Physical Review},
  vol.~106, pp.~893--898, jun 1957.

\bibitem{Dzero2022}
A.~M. Konic, Y.~Zhu, A.~J. Breindel, Y.~Deng, C.~C. Moir, M.~B. Maple, C.~C.
  Almasan, and M.~Dzero, ``Vanishing rkky interactions in ce-based cage
  compounds,'' {\em pre-print arXiv:}, December 2022.

\bibitem{MapleCeAl2}
M.~B. Maple, ``Dependence of $s-f$ exchange on atomic number in rare-earth
  dialuminades,'' {\em Solid State Comm.}, vol.~8, pp.~1915--1917, May 1970.

\bibitem{LPG2005}
V.~Barzykin and L.~P. Gor'kov, ``Competition between phonon superconductivity
  and Kondo screening in mixed valence and heavy fermion compounds,'' {\em
  Phys. Rev. B}, vol.~71, p.~214521, Jun 2005.

\bibitem{ChubukovMaiti}
S.~Maiti and A.~V. Chubukov, ``Superconductivity from repulsive interaction,''
  {\em AIP Conference Proceedings}, vol.~1550, no.~1, pp.~3--73, 2013.

\bibitem{Lavagna1987}
M.~Lavagna, A.~J. Millis, and P.~A. Lee, "d -wave superconductivity in the
  large-degeneracy limit of the Anderson lattice,'' {\em Phys. Rev. Lett.},
  vol.~58, pp.~266--269, Jan 1987.

\bibitem{Read1988}
A.~Houghton, N.~Read, and H.~Won, ``Charge fluctuations, spin fluctuations, and
  superconductivity in the anderson lattice model of heavy-fermion systems,''
  {\em Phys. Rev. B}, vol.~37, pp.~3782--3785, Mar 1988.

\bibitem{Miyake2000}
Y.~Onishi and K.~Miyake, ``Enhanced valence fluctuations caused by f-c Coulomb
  interaction in Ce-based heavy electrons: Possible origin of pressure-induced
  enhancement of superconducting transition temperature in CeCu$_2$Ge$_2$ and
  related compounds,'' {\em Journal of the Physical Society of Japan}, vol.~69,
  no.~12, pp.~3955--3964, 2000.

\bibitem{Coleman2007}
P.~Coleman, {\em Heavy Fermions: Electrons at the Edge of Magnetism}.
\newblock John Wiley \& Sons, Ltd, 2007.

\bibitem{Joerg1}
S.~Fischer, M.~Hecker, M.~Hoyer, and J.~Schmalian, ``Short-distance breakdown
  of the Higgs mechanism and the robustness of the BCS theory for charged
  superconductors,'' {\em Phys. Rev. B}, vol.~97, p.~054510, Feb 2018.

\bibitem{Dzero2016}
M.~Dzero, J.~Xia, V.~Galitski, and P.~Coleman, ``Topological Kondo
  insulators,'' {\em Annual Review of Condensed Matter Physics}, vol.~7, no.~1,
  pp.~249--280, 2016.

\bibitem{Flint2007}
R.~Flint, M.~Dzero, and P.~Coleman, ``Heavy electrons and the symplectic
  symmetry of spin,'' {\em Nat Phys}, vol.~4, pp.~643--648, 08 2008.

\bibitem{AGD}
A.~A. Abrikosov, L.~P. Gorkov, and I.~E. Dzyaloshinski, {\em {\sl Methods of
  Quantum Field Theory in Statistical Physics}}.
\newblock Dover, 1977.

\end{thebibliography}

\end{document}